\documentclass[twocolumn,aps,pra,superscriptaddress,amsmath,floatfix]{revtex4-2}

\usepackage{hyperref}
\hypersetup{
           breaklinks=true,   
           colorlinks=true,   
           pdfusetitle=true,  
           hidelinks=true,
        }

\usepackage{graphicx}
\usepackage{bm}
\usepackage{color}
\usepackage{amssymb}
\usepackage{enumerate}
\usepackage{comment}

\def\la{\langle}
\def\ra{\rangle}

\def\be{\begin{equation}}
\def\ee{\end{equation}}

\begin{document}

\newcommand{\bigjprob}{{\mathcal{P}}}
\newcommand{\bigprob}{_{\bm{q}_F}{\mathcal{P}}_{\bm{q}_I}}
\newcommand{\cum}[1]{\llangle #1 \rrangle}       					
\newcommand{\op}[1]{\hat{\bm #1}}                					
\newcommand{\vop}[1]{\vec{\bm #1}}
\newcommand{\opt}[1]{\hat{\tilde{\bm #1}}}
\newcommand{\vopt}[1]{\vec{\tilde{\bm #1}}}
\newcommand{\td}[1]{\tilde{ #1}}
\newcommand{\mean}[1]{\la#1\ra}                  					
\newcommand{\cmean}[2]{ { }_{#1}\mean{#2}}       				
\newcommand{\pssmean}[1]{ { }_{\bm{q}_F}\mean{#1}_{\bm{q}_I}}
\newcommand{\ket}[1]{\vert#1\ra}                 					
\newcommand{\bra}[1]{\la#1\vert}                 					
\newcommand{\ipr}[2]{\left\la#1\mid#2\right\ra}            				
\newcommand{\opr}[2]{\ket{#1}\bra{#2}}           					
\newcommand{\pr}[1]{\opr{#1}{#1}}                					
\newcommand{\Tr}[1]{\text{Tr}(#1)}               					
\newcommand{\Trd}[1]{\text{Tr}_d(#1)}            					
\newcommand{\Trs}[1]{\text{Tr}_s(#1)}            					
\newcommand{\intd}[1]{\int \! \mathrm{d}#1 \,}
\newcommand{\dd}{\mathrm{d}}
\newcommand{\fullint}{\iint \! \mathcal{D}\mathcal{D} \,}
\newcommand{\drv}[1]{\frac{\delta}{\delta #1}}
\newcommand{\partl}[3]{ \frac{\partial^{#3}#1}{ \partial #2^{#3}} }		
\newcommand{\smpartl}[4]{ \left( \frac{\partial^{#3} #1}{ \partial #2^{#3}} \right)_{#4}}
\newcommand{\smpartlmix}[4]{\left( \frac{\partial^2 #1}{\partial #2 \partial #3 } \right)_{#4}}
\newcommand{\limit}[2]{\underset{#1 \rightarrow #2}{\text{lim}} \;}
\newcommand{\funcd}[2]{\frac{\delta #1}{\delta #2}}
\newcommand{\funcdiva}[3]{\frac{\delta #1[#2]}{\delta #2 (#3)}}
\newcommand{\funcdivb}[4]{\frac{\delta #1 (#2(#3))}{\delta #2 (#4)}}
\newcommand{\funcdivc}[3]{\frac{\delta #1}{\delta #2(#3)}}
\definecolor{dgreen}{RGB}{30,130,30}

\title{Continuous measurement of a qudit using dispersively coupled radiation}
\author{John Steinmetz}
\author{Debmalya Das}
\affiliation{Department of Physics and Astronomy, University of Rochester, Rochester, NY 14627, USA}
\affiliation{Center for Coherence and Quantum Optics, University of Rochester, Rochester, NY 14627, USA}
\affiliation{Institute for Quantum Studies, Chapman University, Orange, CA 92866, USA}
\author{Irfan Siddiqi}
\affiliation{Center for Quantum Coherent Science, Berkeley, CA 94720 USA}
\affiliation{Department of Physics, University of California, Berkeley, CA 94720 USA}
\author{Andrew N. Jordan}
\affiliation{Department of Physics and Astronomy, University of Rochester, Rochester, NY 14627, USA}
\affiliation{Center for Coherence and Quantum Optics, University of Rochester, Rochester, NY 14627, USA}
\affiliation{Institute for Quantum Studies, Chapman University, Orange, CA 92866, USA}

\begin{abstract}
We analyze the continuous monitoring of a qudit coupled to a cavity using both phase-preserving and phase-sensitive amplification. The quantum trajectories of the system are described by a stochastic master equation, for which we derive the appropriate Lindblad operators. The measurement back-action causes spiraling in the state coordinates during collapse, which increases as the system levels become less distinguishable. We discuss two examples: a two-level system and an $N$-dimensional system and meter with rotational symmetry in the quadrature space. We also provide a comparison of the effects of phase-preserving and phase-sensitive detection on the master equation, and show that the average behavior is the same in both cases, but individual trajectories collapse at different rates depending on the measurement axis in the quadrature plane.

\end{abstract}

\maketitle

\section{Introduction}
Continuously monitoring a quantum system allows us to track its behavior throughout the measurement process~\cite{Gisin1992, Barchielli1986, Menskii1998, Caves1986, Caves1987a, Caves1987b, Diosi1988, Barchielli1991, Barchielli1982, Belavkin1992, Gisin1993, Carmichael1993}. Developments in the area of continuous quantum measurement have resulted in many applications for, among other things, quantum information, metrology, and feedback control~\cite{Jacobs2006, Gambetta2008, Barchielli2009, Jacobs2014, Jordan2015, Lewalle2017, Lewalle2020, Wiseman2010, Patti2017, Hacohen2018}. A continuously monitored system is necessarily an open system, since the measurement process requires interaction with the environment~\cite{Haroche2006, Nielsen2010, Wiseman2010, Neumann2018}. To perform each measurement on the system, it is coupled to a meter, and a projective measurement is performed on the meter. When these measurements are performed, the state of the system is disturbed. This disturbance, also known as measurement back-action, can be minimized by weakly coupling the system and meter; this causes a trade-off where very little information is gained from each measurement~\cite{Busch1984, Aharonov1988, Duck1989, Brun2002}.

Continuous monitoring is realized by employing a series of weak measurements in time~\cite{Jacobs2006, Barchielli2009}. The coupling of the qudit with the cavity is dispersive~\cite{Haroche2006}, which means the detuning between the resonator frequency and the system level spacings is much larger than the coupling between the system and the meter. In this regime, the approximate Hamiltonian causes a coherent state of the cavity to undergo a transformation depending on the system state. To continuously track the state of the system, we perform measurements on the cavity field at time intervals that are short compared to the collapse time of the system. Based on the measurement readout and the corresponding inferred states of the system, a stochastic master equation can be obtained which describes the stochastic evolution of the system throughout the measurement process (assuming the dynamics are Markovian). Each solution to the master equation is a diffusive quantum trajectory, which tracks the evolution of the system state in time under both unitary dynamics and measurement back-action. The theory of quantum trajectories has proven to be useful for quantum computing, including having applications for feedback control~\cite{Doherty2000,Wiseman1994,Campagne2013} and entanglement generation~\cite{Ruskov2003,Williams2008,Trauzettel2006,Lewalle2020-limit}, but this has mostly been restricted to trajectories of two-level systems.



In this paper, we study the continuous monitoring of a quantum $N$-level system (or ``qudit") coupled to a microwave cavity, using either phase-sensitive or phase-preserving amplification (corresponding, respectively, to homodyne and heterodyne measurements). It is possible, and sometimes advantageous, to use qudits for computation. They are able to store more information than a qubit and require fewer entangling gates to perform certain algorithms~\cite{Morvan2020,Wang2020,Lanyon2009}. Qutrit-based processors have drawn particular interest, and multiqutrit gate sequences have been implemented using both superconducting and photonic qutrits~\cite{Blok2021,Luo2019}. Higher-dimensional systems are also relevant to standard qubit-based computation since a qubit is generally engineered by using the two lowest energy levels of an anharmonic oscillator as in the case of a superconducting qubit~\cite{Devoret2004}. With an anharmonic potential, the transition frequency between the lowest two levels is sufficiently different from the transitions between the higher-lying states. This enables one to operate within the qubit subspace by using the corresponding transition frequency. In practice, there may be leakage of populations to the higher levels of the system, leading to unwanted signals or errors~\cite{Chen2016, Gambetta2011, Sank2016}. Leakage errors may degrade the performance of error-correcting codes because of a randomizing effect on the qubits affected by leakage~\cite{Fowler2013}. They can also lead to time-correlated errors since such states may be as long-lived as the qubit states, further deteriorating the execution of quantum computing tasks~\cite{Ghosh2013}. The ability to use trajectory-based techniques and to track leakage errors in time provides us with the motivation to study the continuous measurement of a qudit in a microwave cavity.  

At this point it is worthwhile to compare the paradigms of operation of continuous quantum measurements and those in quantum Zeno-like situations~\cite{Misra1977, Nakanishi2001, Facchi2004}. In both cases, the system is measured repeatedly at short time intervals. 
However, measurements leading to the quantum Zeno effect are used to collapse a dynamically evolving quantum state back to its original condition, with vanishing probability for it to wander off. On the other hand, for continuous monitoring, the quantum state changes under the joint influence of both unitary and measurement-induced dynamics. By changing the strength of the measurement, a continuous estimation of the state of the quantum system is possible, even reducing the disturbance of the unitary dynamics to a minimum, at the cost of reducing the acquired information about the system.

There are other types of measurement settings available in the literature, like quantum nondemolition measurements~\cite{Braginskii1967, Braginskii1975, Thorne1978, Braginsky1980, Caves1980, Bocko1982, Brune1992, Jordan2005, Jordan2006b}, for which back-action can be completely evaded. That is, a projective measurement of a single observable can be built up in time. The important criterion for such a measurement is that the observable must commute with the interaction Hamiltonian between the system and meter. This means when a measurement is performed using the meter, there is no back-action on the system, and the measurement result is completely predictable using previous results. For a sequence of measurements of an observable which at different times commutes with itself, one can ensure that there is no back-action. This can be true for the observable at all times or at some regular interval corresponding to continuous and stroboscopic quantum nondemolition measurements, respectively. However, this commutation relation which is so central to the whole scheme is dependent on the free Hamiltonian of the system. This immediately puts a constraint on the types of systems where continuous and stroboscopic quantum nondemolition measurements can be applied.  Also, even if the Hamiltonian is conducive to quantum stroboscopic measurements, the interval between two consecutive stroboscopic measurements cannot be made arbitrarily small because it is constrained by the commutation properties of the relevant observable at different times. In the waiting time between stroboscopic measurements, no knowledge about the trajectory of the qudit is available. It is here that a continuous measurement, approximated as a series of weak measurements at short time intervals, can prove useful.

The measurement process is discussed in depth in Sec.~\ref{sec:meas}. We present a detailed derivation of a stochastic master equation (SME) for phase-preserving amplification, along with the relevant Lindblad operators, and use it to study the behavior of individual stochastic trajectories and ensemble averages in Sec.~\ref{sec:eom}. The equations of motion for the generalized Bloch coordinates are also presented. In Sec.~\ref{sec:examples}, we demonstrate the application of the master-equation approach in order to derive the equations of motion for two examples: a qubit and an $N$-level system with rotational symmetry in the quadrature space. Next, in  Sec.~\ref{sec:comp}, we provide an equivalent derivation using Kraus operators, which allows us to extend the framework to phase-sensitive amplification. While the two measurement readouts of a phase-preserving measurement correspond to the two field quadratures, phase-sensitive measurements give a single readout that corresponds to one field quadrature. Using this approach, the effects of phase-preserving and phase-sensitive amplification are also compared. We conclude in Sec.~\ref{sec:conc} and discuss some possible applications of this work.

\section{Dispersive measurement} \label{sec:meas}

\subsection{Dispersive Jaynes-Cummings model}
Consider a qudit coupled to a resonator, where the qudit interacts with a single field mode. Typically, the Hamiltonian of such a qudit-resonator system would have terms denoting the free Hamiltonians of the qudit and the resonator and a term denoting the interaction between the two. An excitation in the qudit corresponds to the loss of a photon in the resonator, while a relaxation in the qudit creates a photon in the resonator. This situation is best described by the generalized Jaynes-Cummings Hamiltonian
\be 
\begin{split}
H&=\hbar\omega_r \hat{a}^\dagger \hat{a} + \hbar\sum_{j=1}^{N} \omega_j\hat{\Pi}_j \\
&+\hbar\sum_{j=1}^{N-1}g_{j}(\hat{a}^\dagger\hat{\sigma}^-_{j,j+1}+\hat{a}\hat{\sigma}^+_{j,j+1}),
\end{split}
\ee 
where $\hbar\omega_j$ are the qudit-level energies, $\omega_r$ is the resonator frequency, $g_{j}$ is the coupling strength between the resonator and the $j\leftrightarrow j+1$ transition of the qudit, $\hat{\sigma}^{+,-}$ are raising and lowering operators between two adjacent qudit levels, and $\hat{\Pi}_j=\ket{j}\bra{j}$. To read out the qudit state, we work in the dispersive regime, where the detunings $\Delta_j=(\omega_{j+1}-\omega_j)-\omega_r$ between the qudit and resonator field are large compared to the coupling strengths, $4\langle \hat{a}^\dagger \hat{a}\rangle (g_j/\Delta_j)^2\ll 1$. In this limit, we can eliminate the interaction term to lowest order in $\frac{g_j}{\Delta_j}$ by diagonalizing the Hamiltonian~\cite{Blais2020,Koch2007,Boissonneault2010},
\be 
H^D\approx\hbar\omega_r \hat{a}^\dagger \hat{a} +\hbar\sum_{j=1}^{N}(\tilde{\omega}_j+\chi_j\hat{a}^\dagger \hat{a})\hat{\Pi}_j,
\ee 
where $\tilde{\omega}_j=\omega_j+\frac{g_j^2}{\Delta_j}$ are effective frequencies of the qudit levels and $\chi_j=\frac{g^2_{j-1}}{\Delta_{j-1}}-\frac{g^2_j}{\Delta_j}$, with $\chi_1=-\frac{g_1^2}{\Delta_1}$. The frequency of the cavity $\omega_r$ is shifted by a different amount $\chi_j$ depending on the qudit state $\ket{j}$. In this lowest-order dispersive approximation, the cavity frequency shifts linearly with the mean photon number $\la \hat{a}^\dagger \hat{a} \ra$. This approximation breaks down when the power in the cavity becomes comparable to the critical photon number $n_\text{crit}=(\Delta_j/2g_j)^2$, at which point the dependence on the photon number becomes nonlinear~\cite{Gambetta2006,Blais2004}.

If we send in a microwave tone at frequency $\omega_r$, described by a coherent state $\ket{\alpha}$, the qudit-cavity system is initially in the state
\be \label{eq:psi-i}
|\psi(0)\ra = \sum_{j=1}^N c_j\ket{j}\ket{\alpha}.
\ee
During the dispersive interaction, the state evolves according to the unitary operator $e^{-iH^D T/\hbar}$, where $T$ is the length of time for which the qudit and resonator are allowed to interact. Since there is no energy transfer, this results in a state-dependent rotation of the coherent state in the quadrature phase space,
\be
|\psi(T)\ra = e^{-i\tilde{H}_0 T/\hbar}\sum_{j=1}^N c_j \ket{j}\ket{\alpha_j},
\label{eq:ent}
\ee
where $\alpha_j=|\alpha| e^{i\theta_j}$, $\theta_j=-\chi_j T$, and $\tilde{H}_0=\hbar\omega_r \hat{a}^\dagger \hat{a}+\hbar\sum_{j=1}^N \tilde{\omega}_j\hat{\Pi}_j$ is the shifted free Hamiltonian. Throughout this analysis, we will assume that the arrangement operates in the bad-cavity limit $\kappa>>\chi_j$, where $\kappa$ is the cavity line width. Consequently, the light in the resonator escapes from the cavity and is sent to a phase-preserving amplifier that measures the field quadratures. This assumption ensures that the reduced qudit dynamics is Markovian and that the qudit equations of motion in the following sections will not depend on any cavity parameters. 

In what follows, we work in the interaction picture with respect to $\tilde{H}_0$. This is equivalent to transforming to a rotating frame in order to eliminate the dynamics due to $\tilde{H}_0$.

\subsection{Phase-preserving amplification}
In phase-preserving amplification, the cavity field is mixed with a coherent local oscillator, which gives access to both quadratures of the field~\cite{Wiseman2010,Lewalle2020}. This method can be realized by employing parametric amplifiers, traveling-wave parametric amplifiers, and Josephson parametric amplifiers~\cite{Ibarcq2016a, Ibarcq2016b, Caves2012, Reep2019, Yamamoto2008, Bergeal2010}. After mixing with a local oscillator, the two resulting readout signals $I$ and $Q$ are proportional to the real and imaginary parts of a random field amplitude $\beta$~\cite{Jordan2015,Wiseman2010,Carmichael1993}, according to the probability density
\be
P(\beta) = {\rm Tr}_{S,M} [ \hat{\Pi}_\beta |\psi(T)\ra \la \psi(T)| \hat{\Pi}_\beta],
\ee
where $\hat{\Pi}_\beta = |\beta\ra \la \beta|$ and the trace is taken over both the qudit (system) and the resonator field (meter). This is effectively the Husimi $Q$ distribution~\cite{Husimi1940, Sperling2020} of the state $\sum_{j=1}^N |c_j|^2\ket{\alpha_j}\bra{\alpha_j}$. This procedure also causes back-action on the qudit state, altering the coefficients $c_j$. 

We can repeat this procedure many times with a field prepared in the same coherent state $\alpha$. For the moment let us take the case where there is no drive or dephasing, and we delay normalizing the qudit state until the end. After $n$ measurements, returning a set of results $\{\beta_k\}$, the coefficients are updated according to
\be \label{eq:dist-cj}
c'_j = c_j \prod_{k=1}^n  \la \beta_k | \alpha_j\ra,
\ee
where $c_j'$ are unnormalized. Each measurement result is independent of the qudit state at any previous time steps because of the bad-cavity limit. 


\subsection{Time-averaged readout}
The state dynamics do not depend on the order of $\beta_k$, so we can consider the time average of the quadrature results,
\begin{eqnarray}
I &=& \frac{C}{n} \sum_{k=1}^n {\rm Re} \beta_k, \\
Q &=& \frac{C}{n} \sum_{k=1}^n {\rm Im} \beta_k,
\end{eqnarray}
where the constant $C$ is related to the degree of amplification. By the central-limit theorem (which is exact for Gaussian distributions), the probability density for $I$ and $Q$ will also be a mixture of $N$ Gaussians. The means are scaled by the amplification factor $C$, and the variances are reduced by $n$. The photon measurements occur much faster than the system dynamics, so we consider the results to be effectively continuous,
\begin{eqnarray} \label{eq:I-def}
I(t) &=&\frac{C}{\Delta t} \int_{t}^{t+\Delta t} dt' \text{Re}\left[\beta(t')\right], \\
\label{eq:Q-def}
Q(t) &=& \frac{C}{\Delta t} \int_t^{t+\Delta t} dt' \text{Im}\left[\beta(t')\right],
\end{eqnarray}
where $\Delta t$ is the time interval of the measurement.

The probability of obtaining a certain $I$ and $Q$ for the superposition of qudit states is given by
\be \label{eq:dist-P}
P(I,Q) = \sum_j |c_j|^2 P_j(I,Q),
\ee
where
\be \label{eq:dist-Pj}
\begin{split}
P_j(I,Q) \propto \exp\bigg[&- \frac{\Delta t}{4\tau C^2 |\alpha|^2} \big[(I-C|\alpha|\cos\theta_j)^2 \\
&+ (Q-C|\alpha|\sin\theta_j)^2\big]\bigg].
\end{split}
\ee
We use the polar representation of $\alpha$, and $\tau\equiv\frac{T}{4|\alpha|^2}$ is the characteristic measurement time. Each set of results $I,Q$ follows a mixture of $N$ Gaussian probability densities, each associated with a different qudit level. Each Gaussian is centered around the corresponding $\alpha_j$ in both the real and imaginary parts. The time-averaged version of the state disturbance relation \eqref{eq:dist-cj} is
\begin{eqnarray} \label{eq:dist-simple}
c'_j &=& c_j \exp{\left[\frac{\Delta t( I  - i Q)}{4\tau C |\alpha|}e^{i\theta_j}\right]}. \label{dist2}
\end{eqnarray}
The details of this state disturbance calculation can be found in Appendix~\ref{app:K}. Positive $I$ results cause positive phase gain, while the opposite is true for positive $Q$ results. Similarly, depending on the shift in $\alpha$, growing values of $I$ or $Q$ can cause the coefficients of a given state $j$ to be suppressed or enhanced compared to the others, leading to partial collapse of the state.

The full probability density $P(I,Q)$ is a Gaussian with means $\la I \ra = C|\alpha|\sum_j |c_j|^2 \cos\theta_j$ and $\la Q \ra = C|\alpha|\sum_j |c_j|^2 \sin\theta_j$ and variances ${\rm Var}[I] = {\rm Var}[Q] = 2\tau C^2 |\alpha|^2/\Delta t$, with no cross correlation. Consequently, we can write the stochastic signal as
\begin{eqnarray} \label{eq:IQ}
I(t) &=& C|\alpha|\sum_j |c_j|^2 \cos\theta_j +C |\alpha|\sqrt{2\tau} \xi_1(t), \label{noisea} \\
Q(t) &=& C|\alpha|\sum_j |c_j|^2 \sin\theta_j +C |\alpha|\sqrt{2\tau} \xi_2(t), \label{noiseb}
\end{eqnarray}
where $\xi_{1,2}$ are Gaussian white-noise Langevin variables with a mean of zero and variance $\frac{1}{\Delta t}$, obeying
\be
\la \xi_i(t) \xi_j(t') \ra = \delta_{ij} \delta(t-t').
\ee

\subsection{Phase-sensitive amplification}
Instead of performing phase-preserving measurements, one can also carry out phase-sensitive measurements~\cite{Eddins2019, Adamyan2016, Tong2011}. In the phase-sensitive case, one quadrature is amplified while the other is suppressed, which is equivalent to projecting along the eigenstate of one quadrature~\cite{Wiseman2010,Lewalle2020}. This gives one measurement result rather than two, as is typical in homodyne measurements. We measure along the quadrature $\ket{X}$, which is an eigenstate of the dimensionless operator $\hat{X}=\frac{1}{\sqrt{2}}(\hat{a}e^{i\phi}+\hat{a}^\dagger e^{-i\phi})$, where $\phi$ is an angle in the $IQ$ plane measured from the $I$ axis. The $k$th measurement gives a readout $X_k$. The state coefficients are updated according to
\be \label{eq:dist-cj-X}
c_j'=c_j\prod^n_{k=1}\la X_k|\alpha_j\ra,
\ee 
where $\la X_k|\alpha_j\ra$ is a coordinate representation of the coherent state $\ket{\alpha_j e^{-i\phi}}$, which is computed in Appendix~\ref{app:X-alpha}. Similarly to $I$ and $Q$ in the coherent state case, we define the time-averaged readout $r=\frac{C}{n}\sum_{k=1}^n X_k\rightarrow\frac{C}{\Delta t}\int^{\Delta t}_0dt'X(t')$. The total probability density to obtain the readout $r$ is
\be \label{eq:dist-P-r}
P(r)=\sum_{j}|c_j|^2P_j(r),
\ee 
where
\be \label{eq:dist-Pj-r}
P_j(r)\propto\exp\left[-\frac{\Delta t}{4\tau C^2|\alpha|^2}[r-\sqrt{2}C|\alpha|\cos(\theta_j-\phi)]^2\right],
\ee 
similar to \eqref{eq:dist-P} and \eqref{eq:dist-Pj}. The time-averaged state disturbance relation is 
\be \label{eq:dist-simple-r}
c_j'=c_j\exp\bigg[\frac{\Delta t e^{i(\theta_j-\phi)}}{4\tau C|\alpha|}[\sqrt{2}r-C|\alpha|\cos(\theta_j-\phi)]\bigg],
\ee 
where we have followed the same method from Appendix~\ref{app:K}. The total probability density $P(r)$ is a Gaussian with mean $\la r\ra=\sqrt{2}C|\alpha|\sum_{k=1}^n|c_j|^2\cos(\theta_j-\phi)$ and variance $\text{Var}[r]=2\tau C^2 |\alpha|^2/\Delta t$, so we can write the readout as
\be 
r(t)=\sqrt{2}C|\alpha|\sum_{j}|c_j|^2\cos(\theta_j-\phi)+C |\alpha|\sqrt{2\tau}\xi(t).
\label{signal}
\ee 

\section{Equations of motion} \label{sec:eom}
\subsection{Stochastic master equation}
The evolution of the reduced qudit system will be Markovian because we are working in the bad-cavity limit. Since the field escapes the cavity quickly, the equations of motion for the reduced system will not depend on the cavity parameters. The dynamics of a Markovian system being continuously monitored can be described by a stochastic master equation in Lindblad form~\cite{Lindblad1976,Gorini1976,Jacobs2006},
\begin{equation} \label{eq:SME}
\dot{\rho}=-\frac{i}{\hbar}[\hat{H},\rho]+\sum_k \{\mathcal{D}_k[\rho]+\mathcal{M}_k[\rho]\xi_k(t)\}. 
\end{equation}
The sum runs over the measurement operators $\{L_k\}$ applied to the system. The Hamiltonian $\hat{H}$ gives the unitary evolution, the Lindbladian superoperator $\mathcal{D}_k[\rho]=L_k\rho L_k^\dag -\frac{1}{2}(L_k^\dag L_k \rho +\rho L_k^\dag L_k)$ gives the deterministic behavior of the system due to the measurement, and the measurement superoperator $\mathcal{M}_k[\rho]=L_k\rho+\rho L_k^\dag -\rho \text{Tr}(L_k\rho + \rho L_k^\dag)$ gives the stochastic evolution. We have assumed perfect measurement efficiency. The act of measuring the system causes stochastic evolution, which is described by the last term in \eqref{eq:SME}. The stochastic variable $\xi_k=\frac{dW_k(t)}{dt}$ is the $\delta$-correlated Gaussian white noise associated with the $k$th measurement channel, where $W_k(t)$ is a Wiener process.

The derivation of the SME is based on It\^o calculus, which makes the assumptions $dW^2=dt$ and $dt^2=dtdW=0$, where $dW=W(t+dt)-W(t)$ is the Wiener increment~\cite{Jacobs2006,Wiseman2010}. An equation in It\^o form for a density matrix element $\rho_{mn}$ with a set of readouts $\{r_p\}$ can be expressed as
\be \label{eq:Itoform}
d\rho_{mn}=A_{mn}dt+\sum_pB_{mn,p}dW_p.
\ee 
In the following sections, we will derive a stochastic master equation and list the appropriate Lindblad operators for a qudit undergoing both phase-preserving and phase-sensitive measurements.

\subsubsection{Phase-preserving measurement}
We can now derive the equations of motion for a qudit undergoing back-action as a result of phase-preserving measurement. Taking $\Delta t\rightarrow dt$ in the state-update relation \eqref{eq:dist-simple} and inserting the quadrature readouts \eqref{noisea} and \eqref{noiseb}, we obtain It\^o equations for the time-dependent coefficients,
\begin{equation} \label{eq:cjdot}
\begin{split}
dc_j =\bigg[&\sum_k |c_k|^2 e^{-i\theta_k}dt +\sqrt{2\tau}(dW_1-idW_2)\bigg] \frac{e^{i\theta_j}c_j}{4\tau}.
\end{split}
\end{equation}
The density matrix $\rho$ consists of the normalized elements $\rho_{mn}=c_mc_n^*/\sum_k|c_k|^2$. Using \eqref{eq:cjdot} and taking care to apply the rules of It\^o calculus, they are
\be \label{eq:rho-strat}
\begin{split}
\dot{\rho}_{mn}&=-\frac{\rho_{mn}}{4\tau}\left(1-e^{-i\theta_{nm}}\right) \\
&+\frac{\rho_{mn}}{\sqrt{2\tau}}\bigg[\left(\frac{e^{i\theta_m}+e^{-i\theta_n}}{2}-\sum_k\rho_{kk}\cos\theta_k\right)\xi_1 \\
&+\left(\frac{e^{i\theta_m}-e^{-i\theta_n}}{2i}-\sum_k\rho_{kk}\sin\theta_k\right)\xi_2\bigg].
\end{split}
\ee 
These dynamics correspond to a stochastic master equation in the form of \eqref{eq:SME} with $\hat{H}=0$ (due to the use of the interaction picture) and Lindblad operators
\be \label{eq:LI-LQ}
\begin{split}
L_I&=\frac{1}{2\sqrt{2\tau}}\sum_{j}e^{i\theta_j}\hat{\Pi}_j, \\
L_Q&=-\frac{i}{2\sqrt{2\tau}}\sum_{j}e^{i\theta_j}\hat{\Pi}_j.
\end{split}
\ee
The ensemble-averaged behavior is obtained by setting $\xi_1=\xi_2=0$ in \eqref{eq:rho-strat}, which gives a set of differential equations with solutions
\be \label{eq:ens-avg-sol}
\bar{\rho}_{mn}(t)=\bar{\rho}_{mn}(0)e^{-\frac{t}{4\tau}(1-\cos\theta_{nm})-\frac{it}{4\tau}\sin\theta_{nm}},
\ee 
where $\bar{\rho}_{mn}$ are elements of the ensemble-averaged density matrix and $\theta_{nm}\equiv\theta_n-\theta_m$. These dynamics exhibit exponential decay of
the coherences proportional to $(1-\cos\theta_{nm})$ and oscillations proportional to $\sin\theta_{nm}$.

\subsubsection{Phase-sensitive measurement}
The equations of motion of a system undergoing phase-sensitive measurement can be derived in the same way using the state-update relation \eqref{eq:dist-simple-r}. The It\^o equations for the time-dependent coefficients are
\be 
\begin{split}
dc_j=\bigg[& dt\big[2\sum_k|c_k|^2\cos(\theta_k-\phi)-\cos(\theta_j-\phi)\big] \\
&+2\sqrt{\tau}dW \bigg]\frac{e^{i(\theta_j-\phi)}c_j}{4\tau},
\end{split}
\ee 
which lead to equations of motion for the density matrix elements,
\be \label{eq:eom-r}
\begin{split}
\dot{\rho}_{mn}&=-\frac{\rho_{mn}}{4\tau}\left(1-e^{-i\theta_{nm}}\right) \\
&+\frac{\rho_{mn}}{\sqrt{\tau}}\bigg(\frac{e^{i(\theta_m-\phi)}+e^{-i(\theta_n-\phi)}}{2}-\sum_k\rho_{kk}\cos(\theta_k-\phi)\bigg)\xi.
\end{split}
\ee 
Note that this result is the same as \eqref{eq:rho-strat} under the transformation 
\be 
\xi_1\rightarrow\sqrt{2}\xi,\quad
\xi_2\rightarrow 0,\quad \theta_j\rightarrow\theta_j-\phi.
\ee
This can also be summarized by a stochastic master equation with a single Lindblad operator,
\be 
L=\frac{1}{2\sqrt{\tau}}\sum_je^{i(\theta_j-\phi)}\hat{\Pi}_j.
\ee 
Note that this is proportional to the Lindblad operators for phase-preserving measurements in \eqref{eq:LI-LQ} under a change in phase. The total Lindbladian $\sum_k\mathcal{D}_k[\rho]$ is identical to that under phase-preserving measurements, and so the ensemble averages have the same set of solutions \eqref{eq:ens-avg-sol} which are independent of $\phi$. However, the stochastic terms are different, and the two measurement schemes will produce different individual stochastic trajectories. These are known as different ``unravelings" of the same master equation~\cite{Brun2000,Wiseman2010}.

\subsubsection{Inefficient detection}
So far, we have assumed that measurements are done with perfect detector efficiency. In the case of inefficient detection, only a fraction $\eta_k$ of the measurement signal in a channel $k$ is successfully detected. The master equation including inefficient detection is~\cite{Wiseman2010,Jacobs2006}
\begin{equation}
\dot{\rho}=-\frac{i}{\hbar}[\hat{H},\rho]+\sum_k \{\mathcal{D}_k[\rho]+\sqrt{\eta_k}\mathcal{M}_k[\rho]\xi_k(t)\}. 
\end{equation}
The readout signals are altered to 
\begin{eqnarray}
I(t) &=& C|\alpha|\sum_j |c_j|^2 \cos\theta_j +C |\alpha|\sqrt{\tfrac{2\tau}{\eta}} \xi_1(t),  \\
Q(t) &=& C|\alpha|\sum_j |c_j|^2 \sin\theta_j +C |\alpha|\sqrt{\tfrac{2\tau}{\eta}} \xi_2(t)
\end{eqnarray}
for phase-preserving measurements and 
\be 
r(t)=\sqrt{2}C|\alpha|\sum_{j}|c_j|^2\cos(\theta_j-\phi)+C |\alpha|\sqrt{\tfrac{2\tau}{\eta}}\xi(t)
\ee 
for phase-sensitive measurements. The effect of detector inefficiency is to increase the amount of noise in the readout signal.

\subsection{Bloch coordinate dynamics}
The generalized Gell-Mann matrices $\{\Lambda_q\}$~\cite{Bertlmann2008}, an extension of the Pauli matrices, are an appropriate set of basis matrices for an $N$-dimensional system. The density matrix can be expressed~\cite{Hioe1981,Jakobczyk2001,Kimura2003} in terms of these matrices as
\be 
\rho=\frac{1}{N}\mathbb{I}+\frac{1}{2}\mathbf{q}\cdot\mathbf{\Lambda},
\ee 
where $\mathbf{q}$ is an $N$-dimensional generalized Bloch vector. This form is chosen because it preserves the relation $q_i=\text{Tr}(\rho\Lambda_i)$. There are $N^2-1$ generalized Gell-Mann matrices plus the identity, so there are $N^2-1$ Bloch coordinate equations of motion, each given by $\dot{q}=\text{Tr}(\Lambda_q\dot{\rho})$. There are $N-1$ diagonal generalized Gell-Mann matrices, and as shown in \eqref{eq:rho-strat}, the diagonal elements $\rho_{nn}$ all have only stochastic evolution and no drift. The remaining matrices take the form $\Lambda_{mn}=\ket{\psi_m}\bra{\psi_n}+\ket{\psi_n}\bra{\psi_m}$ for $m<n$ and $\Lambda_{mn}=-i(\ket{\psi_n}\bra{\psi_m}-\ket{\psi_m}\bra{\psi_n})$ for $m>n$. These result in the equations of motion
\be 
\begin{split}
\dot{q}_{mn}&=\dot{\rho}_{mn}+\dot{\rho}_{nm},\quad m<n, \\
\dot{q}_{mn}&=-i(\dot{\rho}_{mn}-\dot{\rho}_{nm}),\quad m>n.
\end{split}
\ee 
Here we will focus on the Bloch coordinate dynamics of a qudit undergoing phase-preserving measurement, but the phase-sensitive case can be studied in the same way. Inserting expression \eqref{eq:rho-strat}, we get the coupled pairs of equations in It\^o form,
\be \label{eq:qdot-ij}
\begin{split}
\dot{q}_{mn}&=-\frac{q_{mn}}{4\tau}(1-\cos\theta_{nm})-\frac{q_{nm}}{4\tau}\sin\theta_{nm} \\
&\quad+\frac{1}{2\sqrt{2\tau}}\bigg[q_{mn}(\xi_m+\xi_n-2\sum_k\rho_{kk}\xi_k)+q_{nm}(\xi_m'-\xi_n')\bigg] \\
\dot{q}_{nm}&=-\frac{q_{nm}}{4\tau}(1-\cos\theta_{nm})+\frac{q_{mn}}{4\tau}\sin\theta_{nm} \\
&\quad+\frac{1}{2\sqrt{2\tau}}\bigg[q_{nm}(\xi_m+\xi_n-2\sum_k\rho_{kk}\xi_k)-q_{mn}(\xi_m'-\xi_n')\bigg],
\end{split}
\ee 
where $\xi_m=\xi_1\cos\theta_m+\xi_2\sin\theta_m$, $\xi_m'=\xi_1\sin\theta_m-\xi_2\cos\theta_m$, and $m<n$. The drift terms are the equations for a damped harmonic oscillator, which have the solutions
\be
\begin{split}
\bar{q}_{mn}(t)&=e^{-\gamma_{nm} t}\left[\bar{q}_{mn}(0)\cos\omega_{nm} t-\bar{q}_{nm}(0)\sin\omega_{nm} t\right] \\
\bar{q}_{nm}(t)&=e^{-\gamma_{nm} t}\left[\bar{q}_{nm}(0)\cos\omega_{nm} t+\bar{q}_{mn}(0)\sin\omega_{nm} t\right],
\end{split}
\ee 
where $\omega_{nm}=\frac{\sin\theta_{nm}}{4\tau}$, $\gamma_{nm}=\frac{1-\cos\theta_{nm}}{4\tau}$, and $\bar{q}_{mn}$ are ensemble-averaged Bloch coordinates. This means that there is spiraling in each pair of coordinates $(\bar{q}_{mn},\bar{q}_{nm})$ that is independent of all the other coordinates. 

The diagonal coordinates $\bar{q}_{nn}$ (where $0\leq n\leq N-1$) have only stochastic evolution and no drift. These coordinates can be written in terms of the diagonal density-matrix elements as
\be 
\bar{q}_{nn}=\sqrt{\frac{2}{(n+1)(n+2)}}\sum_{k=0}^{n}\bar{\rho}_{kk}-\sqrt{\frac{2(n+1)}{n+2}}\bar{\rho}_{n+1,n+1}.
\ee
We can insert the noise terms from \eqref{eq:rho-strat} into the time derivative of the above expression to obtain the stochastic evolution.

\section{Examples} \label{sec:examples}
\subsection{$N$-level clock system}
\begin{figure}
    \centering
    \includegraphics[width=0.6\linewidth]{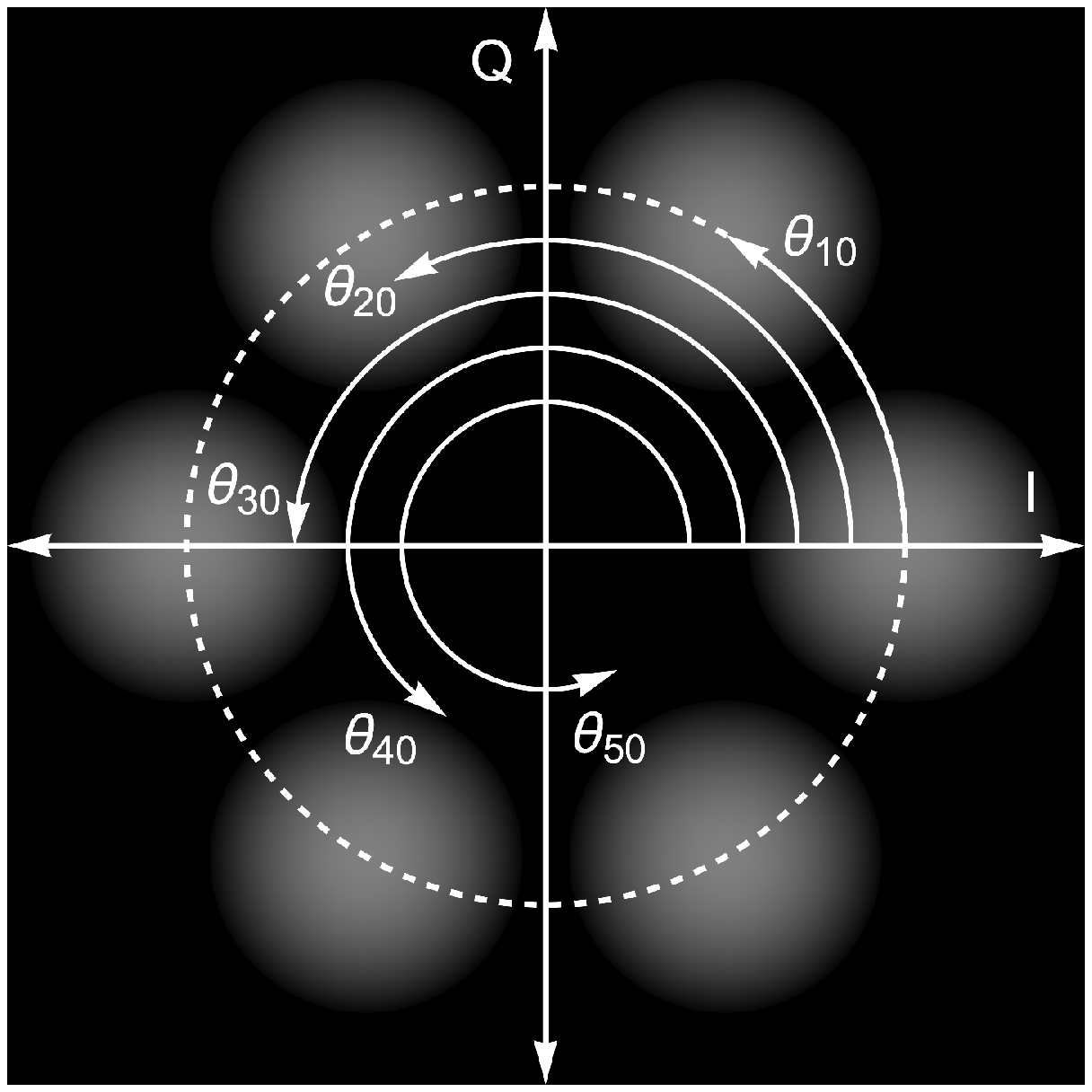}
    
    \vspace{20pt}
    
    \includegraphics[width=\linewidth]{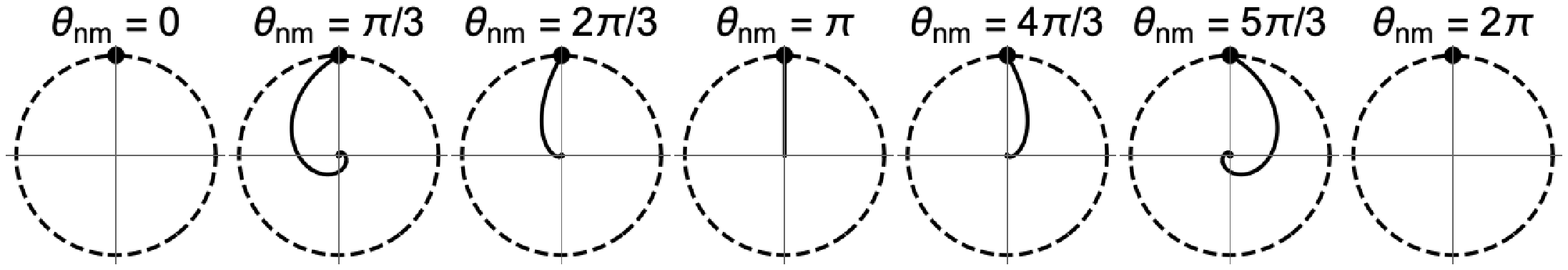}
    \begin{picture}(0,0)
    \put(-67,208){\textcolor{white}{a)}}
    \put(-115,60){\textcolor{black}{b)}}
    \end{picture}
    \caption{(a) An illustration of the probability densities associated with a six-level clock system, where $\theta_{j0}\equiv\theta_{j}=\frac{\pi j}{3}$, in the $IQ$ plane. The circular arrows indicate the angles $\theta_1,\ldots,\theta_5$. (b) Pairs of spiraling coordinates for initial conditions $\bar{q}_{mn}(0)=0$ and $\bar{q}_{nm}(0)=1$. Each curve is plotted on a unit circle in the $(\bar{q}_{nm},\bar{q}_{mn})$ plane. Since $\theta_{nm}$ sets the frequency and decay constant, pairs of coordinates with the same angular separation $\theta_{nm}$ have the same behavior, so although there are $35$ coordinates in a six-level system, these plots show the only possible spirals. Each plot corresponds to coordinates appearing in a specific super- or subdiagonal of $\bar{\rho}$; the first and last plots, where the state does not evolve, correspond to coordinates appearing on the main diagonal.}
    \label{fig:clock}
\end{figure}

An $N$-level ``clock" system, as illustrated in Fig.~\ref{fig:clock}(a), is one in which the quadrature angles $\theta_j$ are distributed evenly at intervals of $2\pi/N$. In this case, the Lindblad operators are $L_I=\frac{1}{2\sqrt{2\tau}}\sum_j e^{-i 2\pi j/N}\hat{\Pi}_j$ and $L_Q=-\frac{i}{2\sqrt{2\tau}}\sum_j e^{-i 2\pi j/N}\hat{\Pi}_j$. This results in the ensemble-averaged dynamics
\be
\begin{split}
\bar{q}_{mn}(t)&=e^{-\gamma_{nm} t}\left[\bar{q}_{mn}(0)\cos\omega_{nm} t-\bar{q}_{nm}(0)\sin\omega_{nm} t\right], \\
\bar{q}_{nm}(t)&=e^{-\gamma_{nm} t}\left[\bar{q}_{nm}(0)\cos\omega_{nm} t+\bar{q}_{mn}(0)\sin\omega_{nm} t\right],
\end{split}
\ee 
where $\omega_{nm}=\frac{\sin[\frac{2\pi}{N}(n-m)]}{4\tau}$ and $\gamma_{nm}=\frac{1-\cos[\frac{2\pi}{N}(n-m)]}{4\tau}$. As in the general case, the coordinates $\bar{q}_{nn}$ corresponding to diagonal generalized Gell-Mann matrices are unaffected. See Fig.~\ref{fig:clock}(b) for plots of the spiraling Bloch coordinates in a six-level clock system. The frequency and decay rate of the spirals depend only on how far away from the diagonal the pair of coordinates appears in the density matrix. Any pairs of coordinates $\bar{q}_{mn}$ and $\bar{q}_{nm}$ that appear in the same super- or subdiagonal of $\bar{\rho}$ will exhibit the same dynamics. Higher-dimensional systems have the potential for smaller angle differences $\theta_{nm}$ since there are more levels to distribute around the circle, and therefore they have the potential for slower spirals. Fig.~\ref{fig:clock} shows the range of spirals that occur in a six-level system. All the simulations here are done assuming the use of the interaction picture.

\subsection{Qubit spiral trajectories}
\begin{figure}
    \centering
    
    \begin{picture}(170,170)
        \put(-35,-10){\rotatebox{90}{\includegraphics[scale=0.5,trim={120 150 140 160},clip]{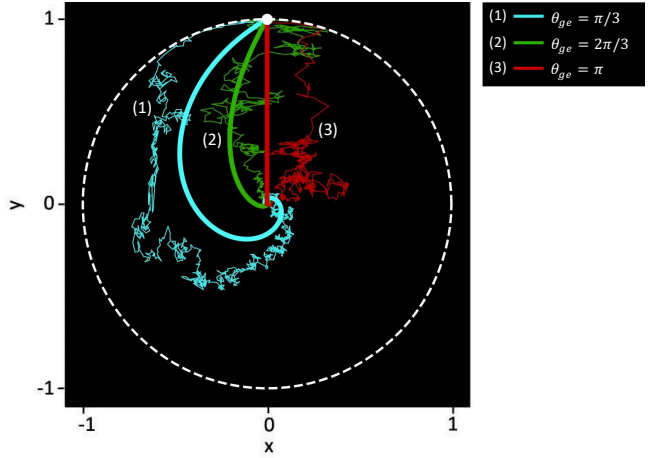}}}
    \end{picture}

    \caption{Sample dynamics in the phase-preserving amplification case for a qubit in the $xy$ plane initially at $\mathbf{q}=(0,1,0)$ undergoing measurement back-action with $\tau=1~\mu\text{s}$ and $dt=10~\text{ns}$ for different values of $\theta_{ge}\equiv\theta_g-\theta_e$. The simulation was run for $50~\mu\text{s}$ to allow time for each state to decohere completely. Each smooth curve is an ensemble average of trajectories, derived using the drift term of the master equation. One sample trajectory is shown for each value of $\theta_{ge}$. When $\theta_{ge}=\pi$, the averaged dynamics decay exponentially without any spiraling. As $\theta_{ge}\rightarrow 0$, the probability densities get closer together, so there is less decoherence; the spiral approaches a unit circle, and the period of rotation gets longer. When $\theta_{ge}=0$, in which case the two measurement outcomes are equivalent, the spiral is simply the unit circle, but the frequency is $\omega=0$, so the state is unaffected. This case of indistinguishable measurement outcomes is equivalent to not making any measurement. The region $\theta_{ge}\in[\pi,2\pi]$ has the same dynamics, but the rotations are clockwise.}
    \label{fig:spiral}
\end{figure}

We can also use this framework to recover the behavior of a qubit under both measurement schemes. As an example, we present the results for phase-preserving measurements. The deterministic equations of motion, derived by averaging over the noise terms in the SME, are
\be 
\begin{split}
\dot{\bar{x}}&=-\frac{\bar{x}}{4\tau}(1-\cos\theta_{ge})-\frac{\bar{y}}{4\tau}\sin\theta_{ge}, \\
\dot{\bar{y}}&=-\frac{\bar{y}}{4\tau}(1-\cos\theta_{ge})+\frac{\bar{x}}{4\tau}\sin\theta_{ge}, \\
\dot{\bar{z}}&=0,
\end{split}
\ee 
which have the deterministic solutions
\be 
\begin{split}
\bar{x}(t)&=e^{-\gamma_{ge} t}\left[\bar{x}(0)\cos\omega_{ge} t-\bar{y}(0)\sin\omega_{ge} t\right], \\
\bar{y}(t)&=e^{-\gamma_{ge} t}\left[\bar{y}(0)\cos\omega_{ge} t+\bar{x}(0)\sin\omega_{ge} t\right], \\
\bar{z}(t)&=\bar{z}(0),
\end{split}
\ee 
where $\omega_{ge}=\frac{\sin\theta_{ge}}{4\tau}$ and $\gamma_{ge}=\frac{1-\cos\theta_{ge}}{4\tau}$. We have chosen $\theta_0\rightarrow\theta_e$ and $\theta_1\rightarrow\theta_g$ to be the dispersive shifts for the readout corresponding, respectively, to the excited and ground states of the qubit. If we choose a separation along the $I$ quadrature, i.e., $\theta_{ge}=\pi$, this reduces to the standard exponential decay along $x$ and $y$~\cite{Wiseman2010,Murch2013}. For all other values of $\theta_{ge}$, there is spiraling in the $xy$ plane. The frequency of the spiraling is largest when $\theta_{ge}=\pi/2$. If $\theta_{ge}=0$, the two measurement outcomes are indistinguishable, and there is no measurement back-action at all. The quadrature angles $\theta_j$, as shown in \eqref{eq:ent}, are set by the dispersive shifts $\chi_j$ and the qubit-resonator interaction time $T$. Some examples of these dynamics are shown in Fig.~\ref{fig:spiral}.

\section{Comparison of phase-preserving and phase-sensitive measurement} \label{sec:comp}

\subsection{Stochastic quantum trajectories}
\begin{figure}
    \centering
    \begin{picture}(100,140)
        \put(-80,-10){\includegraphics[scale=0.68]{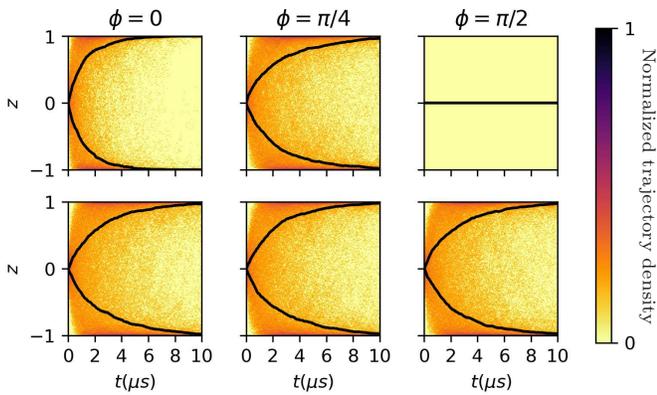}}
        \put(165,125){\rotatebox{270}{\scriptsize{Normalized trajectory density}}}
    \end{picture}
   \caption{Distribution of $1000$ trajectories of the $z$ coordinate of a qubit according to \eqref{eq:traj-p} and \eqref{eq:traj-s}. The qubit is initialized at $\mathbf{q}=(0,1,0)$ and measured using phase-sensitive (top row) and phase-preserving (bottom row) amplification, with dispersive shifts $\theta_e=0$ and $\theta_g=\pi$ and time-scales $\tau=1~\mu\text{s}$ and $dt=50~\text{ns}$. Each column corresponds to a different amplification angle $\phi$. Ensemble averages (black) are shown for trajectories postselected on $z\geq0$ and $z<0$. The normalized density of trajectories is shown on the color bar. The phase-sensitive measurement decays slower as $\phi\rightarrow\pi/2$ since the amplification axis gets farther from the informational quadrature. The phase-preserving measurement is unaffected by changes in $\phi$ and decays at the same rate as the phase-sensitive measurement for $\pi/4$.}
    \label{fig:comp}
\end{figure}

\begin{figure}
    \centering
    \includegraphics[scale=0.4]{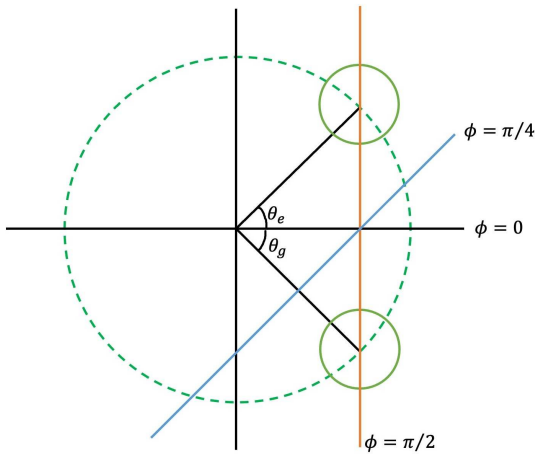}
    \caption{Different orientations $\phi$ of the phase-sensitive measurement axis in the $IQ$ plane for $N=2$. The two small circles are the locations $\theta_e$ and $\theta_g$ of the probability densities associated with the two eigenstates. In general, the orientation that gives the best discrimination is $\phi_{opt}=\frac{\pi}{2}-\frac{\theta_e+\theta_g}{2}$.}
    \label{fig:orient}
\end{figure}

Although the Lindbladian (nonstochastic) master equations are identical for phase-preserving and phase-sensitive measurements and are unaffected by changes in $\phi$, the stochastic trajectories can behave differently. This is an example of quantum steering~\cite{Schrodinger1935,Schrodinger1936,Wiseman2007}, where by changing the axis along which a system (in this case the cavity field) is measured, we can alter the behavior of an entangled system (the qudit). As an example, we study the rates at which a state collapses under these different types of measurements. These dynamics are shown in Fig.~\ref{fig:comp}. If we set the dispersive shifts to be $\theta_e=0$ and $\theta_g=\pi$ but amplify along an axis $\ket{X}$, the noise term in $\dot{z}$ is
\be \label{eq:traj-p}
\dot{z}=\frac{1}{\sqrt{2\tau}}(1-z^2)(\xi_1\cos\phi-\xi_2\sin\phi)
\ee 
for phase-preserving measurements (using the same $\theta_j\rightarrow\theta_j-\phi$ transformation) and
\be \label{eq:traj-s}
\dot{z}=\frac{\xi}{\sqrt{\tau}}(1-z^2)\cos\phi
\ee 
for phase-sensitive measurements. Compare $(\xi_1\cos\phi-\xi_2\sin\phi)/\sqrt{2}$, which has variance $\frac{1}{2\Delta t}$, to $\xi\cos\phi$, which has variance $\frac{\cos^2\phi}{\Delta t}$. Note that the variance caused by phase-preserving measurements is independent of the axis $\phi$. When $\phi=0$ (left column of Fig.~\ref{fig:comp}), phase-sensitive measurements cause the state to collapse more quickly than phase-preserving measurements since they are exclusively measured along the informational quadrature, while phase-preserving measurements are also measured along the quadrature containing no information. When $\phi=\pi/4$ (middle column of Fig.~\ref{fig:comp}), the noise has the same variance for both types of measurements, so the states collapse at the same rate. When $\phi=\pi/2$ (right column of Fig.~\ref{fig:comp}), phase-sensitive measurements detect only noise and give no information at all about the system state, so the state does not collapse. This is an example of a no-knowledge measurement, which has applications in feedback control~\cite{Szigeti2014,Saiphet2020}. When using phase-sensitive measurements on a qubit, the optimal value of $\phi$ is $\phi_{opt}=\frac{\pi}{2}-\frac{\theta_e+\theta_g}{2}$, as shown in Fig.~\ref{fig:orient}.

\subsection{Distinguishability of qudit states}
The collapse rate of a qudit due to measurement back-action can be predicted by using the signal-to-noise ratio of the measurement. We can quantify this by looking at the distinguishability between pairs of Gaussian probability densities $P_i$ and $P_j$, corresponding to distinguishing qudit state $i$ from $j$. We quantify the signal $S$ for phase-preserving measurements using the Bhattacharyya distance~\cite{Bhattacharyya1946,Fukunaga2013}
\be \label{eq:Bhatt}
S_{ij}=-\ln\left({\int dI dQ \sqrt{P_i P_j}}\right),
\ee 
which is a measure of the distinguishability of two probability densities. The integral is over the two measurement quadratures $I$ and $Q$, defined in \eqref{eq:I-def} and \eqref{eq:Q-def}. For phase-preserving measurements, we use $P_i(I,Q)$ and $P_j(I,Q)$, given in \eqref{eq:dist-Pj}. The signal after a time $\Delta t$ is
\be
S_{ij}=\frac{\Delta t}{4T}|\alpha_{i}-\alpha_{j}|^2,
\ee
which is the distance between the probability densities on the $IQ$ plane scaled by the inverse of their width. 

For phase-sensitive measurements, we use the probability densities $P_i(r)$ and $P_j(r)$ given in \eqref{eq:dist-Pj-r} and use the rotated basis $\theta_j\rightarrow\theta_j-\phi$. This gives
\be
P_j(r)=\sqrt{\frac{\Delta t}{4\pi\tau C^2 |\alpha|^2}}e^{-\frac{\Delta t}{4\tau C^2 |\alpha|^2}(r-\sqrt{2}C|\alpha|\cos(\theta_j-\phi))^2}.
\ee
Using a single-variable version of \eqref{eq:Bhatt}, the signal obtained using phase-sensitive measurements is
\be 
S_{ij}=\frac{\Delta t|\alpha|^2}{2T}\left[\cos(\theta_i-\phi)-\cos(\theta_j-\phi)\right]^2.
\ee
Take, for example, the qubit system in Fig.~\ref{fig:comp}, where $\theta_e=0$ and $\theta_g=\pi$. Phase-preserving measurements give a signal of $S_{ij}=\frac{\Delta t |\alpha|^2}{T}$, which is independent of the measurement axis $\phi$. Phase-sensitive measurements give a signal of $S_{ij}=\frac{2\Delta t |\alpha|^2}{T}\cos^2\phi$. The signals are equal when $\phi=\frac{\pi}{4}$, just as the variances of trajectories are equal in Fig.~\ref{fig:comp}. When we measure along the informational quadrature $\phi=0$, phase-sensitive measurements give more information about the system than phase-preserving measurements since none of the readout is wasted on the no-information quadrature. When we measure along the no-information quadrature $\phi=\pi/2$, phase-sensitive measurements give no information, i.e., $S_{ij}=0$.

\section{Conclusion}
\label{sec:conc}

We have modeled the continuous monitoring of a qudit using phase-preserving and phase-sensitive measurements using a stochastic master equation and identified the appropriate Lindblad operators. The SME was used to analyze two example systems: a qubit and an $N$-level system with discrete rotational symmetry. We have shown that phase-preserving and phase-sensitive measurements have the same effect on the average dynamics but cause individual stochastic quantum trajectories to behave differently, such as by inducing different collapse rates. In the qubit case, the measurement results in pairs of ensemble-averaged Bloch coordinates behave like a damped harmonic oscillator, exhibiting the usual decoherence but also spiraling around the phase space. The spiraling is stronger when the dispersive shifts are smaller and completely absent when they lie at diametrically opposite points in the $IQ$ plane.

This theoretical framework can be used to model and simulate stochastic trajectories for higher-dimensional systems such as qutrits. This allows us to study the behavior of qudits undergoing any arbitrary operations and could lead to applications for feedback control. In addition, the physical qubits used for quantum computing are often made using a two-level subspace in a higher-dimensional system. While the dynamics are mostly contained to the qubit subspace, there are often leakage errors in higher levels, which can be tracked in time by making continuous weak measurements and treating the system as a qutrit or higher-dimensional system.

\section{Acknowledgments}
This work is supported by U.S. Army Research Office grant no. W911NF-18-10178 and NSF grant no. DMR-1809343. We thank G. Koolstra, P. Lewalle, W. P. Livingston, R. Naik, D. I. Santiago, K. Siva, and N. J. Stevenson for many helpful discussions.

\appendix
\begin{widetext}
\section{State disturbance calculation} \label{app:K}
Starting from Eq.~\eqref{eq:dist-cj},
\be 
\begin{split}
c_j'&= c_j \prod_{k=1}^n  \la \beta_k | \alpha_j\ra \\
&=c_j\exp\left[\sum_{k=1}^n-|\beta_k-\alpha_j|^2/2+i\text{Im}(\beta_k^*\alpha_j)\right] \\
&=c_j\exp\bigg[\sum_{k=1}^n-\frac{(\text{Re}\beta_k-\text{Re}\alpha_j)^2}{2}\bigg]\exp\bigg[\sum_{k=1}^n-\frac{(\text{Im}\beta_k-\text{Im}\alpha_j)^2}{2}\bigg]\exp\bigg[\sum_{k=1}^ni(\text{Re}\beta_k\text{Im}\alpha_j-\text{Im}\beta_k\text{Re}\alpha_j)\bigg].
\end{split}
\ee 

The product of $n$ Gaussian functions is a Gaussian with variance $\frac{1}{\sigma^2}=\sum_{k=1}^n\frac{1}{\sigma_k^2}$ and mean $\mu=\sigma^2\sum_{k=1}^n\frac{\mu_k}{\sigma_k^2}$. Applying these rules and changing the sums to integrals using $\frac{1}{n}\sum_{k=1}^n\rightarrow\frac{1}{\Delta t}\int_0^{\Delta t} dt$ give
\be 
\begin{split}
c_j'&=c_j\exp\bigg[-\frac{n}{2C^2}(I-C|\alpha|\cos\theta_j)^2\bigg]\exp\bigg[-\frac{n}{2C^2}(Q-C|\alpha|\sin\theta_j)^2\bigg]\exp\bigg[\frac{in|\alpha|}{C}(I\sin\theta_j-Q \cos\theta_j)\bigg].
\end{split}
\ee 
Defining the characteristic measurement time $\tau\equiv\frac{T}{4|\alpha|^2}$, where $T=\frac{\Delta t}{n}\sim\frac{1}{\kappa}$ is the time each photon has to interact with the qudit, we get
\be 
c_j'=c_j\exp\bigg[-\frac{\Delta t}{8\tau C^2 |\alpha|^2}(I-C|\alpha|\cos\theta_j)^2\bigg]\exp\bigg[-\frac{\Delta t}{8\tau C^2|\alpha|^2}(Q-C|\alpha|\sin\theta_j)^2\bigg]\exp\bigg[\frac{i\Delta t}{4\tau C|\alpha|}(I\sin\theta_j-Q\cos\theta_j)\bigg].
\ee 
We can also drop any term in the exponent that is not state dependent and the terms quadratic in the time-averaged quadratures, because they drop out when the state is normalized. The terms $|\alpha_j|^2$ are the same for all $j$ because of the dispersive approximation, so they also drop out. This leaves a simplified update equation,
\begin{eqnarray}
c'_j &=& c_j \exp{\left[\frac{\Delta t( I  - i Q)}{4\tau C |\alpha|}e^{i\theta_j}\right]},
\end{eqnarray}
which matches \eqref{eq:dist-simple}. Using the same method, it can be shown that for phase-sensitive measurements, the state-update relation \eqref{eq:dist-cj-X} becomes
\be 
\begin{split}
c_j'&=c_j\prod^n_{k=1}\la X_k|\alpha_j\ra \\
&=c_j\exp\bigg[\sum^n_{k=1}\bigg(-\frac{1}{2}[X_k-\sqrt{2}\alpha\cos(\theta_j-\phi)]^2+i\text{Im}\alpha_j[\sqrt{2}X_k-\alpha\cos(\theta_j-\phi)]\bigg)\bigg] \\
&=c_j\exp\bigg[-\frac{\Delta t}{8\tau C^2|\alpha|^2}[r-\sqrt{2}C|\alpha|\cos(\theta_j-\phi)]^2\bigg]\exp\bigg[\frac{i\Delta t \sin(\theta_j-\phi)}{4\tau C |\alpha|}[\sqrt{2}r-C|\alpha|\cos(\theta_j-\phi)]\bigg],
\end{split}
\ee 
where $\la X_k | \alpha_j \ra$ is computed in Appendix~\ref{app:X-alpha}. By dropping terms that will cancel out when the state is normalized, this can be simplified to 
\be
c_j'=c_j\exp\bigg[\frac{\Delta t e^{i(\theta_j-\phi)}}{4\tau C|\alpha|}[\sqrt{2}r-C|\alpha|\cos(\theta_j-\phi)]\bigg],
\ee 
which matches \eqref{eq:dist-simple-r}.

\end{widetext}

\section{Coordinate representation of a coherent state} \label{app:X-alpha}
To derive the coordinate representation $\la X|\alpha\ra$ of a coherent state, where $X$ is dimensionless, we use the Fock basis representation
\be 
\ket{\alpha}=e^{-\frac{|\alpha|^2}{2}}\sum_{n=0}^\infty\frac{\alpha^n}{\sqrt{n!}}\ket{n}
\ee
to write
\be 
\la X|\alpha\ra =e^{-\frac{|\alpha|^2}{2}}\sum_{n=0}^\infty\frac{\alpha^n}{\sqrt{n!}}\psi_n,
\ee
where $\psi_n$ are the quantum harmonic oscillator eigenstate wave functions. Inserting these gives
\be 
\la X|\alpha\ra =e^{-\frac{|\alpha|^2}{2}}\sum_{n=0}^\infty\frac{\alpha^n}{\sqrt{2^n}n!}e^{-\frac{X^2}{2}}H_n(X),
\ee
where $H_n$ are the Hermite polynomials. Using the generating function 
\be 
e^{2Xt-t^2}=\sum_{n=0}^\infty H_n(X)\frac{t^n}{n!}
\ee 
for the Hermite polynomials, this becomes
\be
\begin{split}
\la X|\alpha\ra &=\exp\bigg[-\frac{|\alpha|^2}{2}-\frac{X^2}{2}+\sqrt{2}X\alpha-\frac{\alpha^2}{2}\bigg] \\
&=\exp\bigg[-\frac{1}{2}[X-\sqrt{2}\text{Re}(\alpha)]^2 \\
&\quad\quad\quad+i\text{Im}(\alpha)[\sqrt{2}X-\text{Re}(\alpha)]\bigg].
\end{split}
\ee

\section{Kraus operator derivation of the equations of motion} \label{app:kraus}
\subsection{Kraus operator description of phase-preserving measurement}
Here we perform a calculation equivalent to the one in Sec.~\ref{sec:eom} using Kraus operators instead of a stochastic master equation~\cite{Lewalle2020}. As mentioned previously, making a phase-preserving measurement on the joint qudit-resonator state~\eqref{eq:ent} gives the real and imaginary parts of a field amplitude $\beta$. We define the coherent-state phase-shift operator~\cite{Leonhardt1997}
\be
\hat{U}(\theta_j)=e^{i\theta_j \hat{a}^\dagger \hat{a}},
\ee
which preserves the width of a coherent state $\ket{\alpha}$ but causes a rotation $\ket{\alpha}\rightarrow \ket{\alpha_j}\equiv\ket{\alpha e^{i\theta_j}}$ in the quadrature phase space. We can then write a Kraus operator
\be
\begin{split}
M_\beta&=\bra{\beta}\left(\sum_j\hat{\Pi}_j\hat{U}(\Delta_j)\right)\ket{\alpha} \\
&= \sum_j\la\beta|\alpha_j\ra\hat{\Pi}_j
\end{split}
\ee 
that applies the measurement to the field mode and acts only on the qudit state. For a series of $n$ measurements, this becomes
\be 
M_\beta=\prod_{k=1}^n\left(\sum_j\la\beta_k|\alpha_j\ra\hat{\Pi}_j\right).
\ee 
This Kraus operator is normalized such that $\frac{dt}{4\pi\tau C^2 |\alpha|^2}\int\int_{-\infty}^\infty dI dQ M_\beta M_\beta^\dagger=\mathbb{I}$. This operator can be approximated as $M_\beta\approx\mathbb{I}+(\lambda\mathbb{I}+m_\beta) dt$, where $\lambda$ is a constant and
\be 
m_\beta=\frac{I-iQ}{4\tau C |\alpha|^2}
\sum_j\alpha_j\hat{\Pi}_j.
\ee 
The equations of motion that result from applying this Kraus operator to the system are given~\cite{Lewalle2020} by
\be \label{eq:eom-kraus}
\dot{\rho}\approx m_\beta\rho+\rho m_\beta^\dagger -\rho\text{Tr}(m_\beta\rho+\rho m_\beta^\dagger).
\ee 
Once the readout signals \eqref{eq:IQ} are substituted in the equation, the resulting expressions will take the form
\be \label{eq:form-strat}
d\rho_{mn}=\tilde{A}_{mn}dt+\sum_pB_{mn,p}dW_p,
\ee 
where $p=1,2$ correspond to $I,Q$. It should be noted that these equations of motion do not match the corresponding equations \eqref{eq:rho-strat} that were derived using a stochastic master equation, which have the form
\be \label{eq:form-ito}
d\rho_{mn}=A_{mn}dt+\sum_pB_{mn,p}dW_p.
\ee 
However, we can transform between them using
\be \label{eq:strat-ito}
A_{mn} = \tilde{A}_{mn} + \frac{1}{2}\sum_{i,j}\sum_{p} B_{ij,p} \partial_{ij} B_{mn,p},
\ee
where $\partial_{ij}\equiv\frac{\partial}{\partial\rho_{ij}}$. The sum over $(m,n)$ excludes the final element $\rho_{N,N}$ since it can be expressed in terms of the other elements using $\text{Tr}(\rho)=1$. This is the standard transformation for converting between It\^o and Stratonovich equations~\cite{Jacobs2006,Wiseman2010,Gough2018}. When \eqref{eq:form-strat} is subjected to a Stratonovich integral, it will give the same result as when \eqref{eq:form-ito} is subjected to an It\^o integral.


\subsection{Phase-sensitive measurement}
Now we will look at phase-sensitive measurements along the quadrature $\ket{X}$, which is an eigenstate of the dimensionless operator $\hat{X}=\frac{1}{\sqrt{2}}(\hat{a}e^{i\phi}+\hat{a}^\dagger e^{-i\phi})$. We can write the Kraus operator
\be 
M_X=\sum_j\la X|\alpha_j\ra\hat{\Pi}_j,
\ee 
where the coefficients are the coordinate representation of the coherent state $\ket{\alpha_j}$ (see Appendix~\ref{app:X-alpha} for details). For $n$ measurements, the corresponding Kraus operator is given by
\be 
M_X=\prod_{k=1}^n\left(\sum_j\la X_k|\alpha_j\ra\hat{\Pi}_j\right).
\ee 
Like in the phase-preserving case, we expand $M_X$ in $\Delta t$ to obtain the first-order term
\be 
m_x=\frac{1}{4\tau C |\alpha|}\sum_j e^{i\theta_j}(\sqrt{2}r-C|\alpha| \cos\theta_j )\hat{\Pi}_j,
\ee 
which gives the Stratonovich equations of motion by using \eqref{eq:eom-kraus} and is related to the It\^o equation \eqref{eq:eom-r} by the transformation \eqref{eq:strat-ito}. The drift terms are unchanged from the phase-preserving case, but the noise terms have undergone the transformations $\xi_1\rightarrow\sqrt{2}\xi$ and $\xi_2\rightarrow 0$, effectively concentrating the noise into a single quadrature. To measure along an arbitrary quadrature $\ket{X}$, we substitute $a\rightarrow a e^{i\phi}$, which implies $\theta_j\rightarrow\theta_j-\phi$.

\bigskip
\bibliography{refs}

\end{document}